\documentclass[aps,pre,reprint,groupedaddress, showpacs]{revtex4-1}

\usepackage{float}
\usepackage[overload]{textcase}

\usepackage{graphicx}	

\usepackage[force]{feynmp-auto}
\usepackage{enumitem}
\usepackage{amssymb}
\usepackage{amsmath}
\usepackage{mathtools}
\usepackage{MnSymbol}
\usepackage{esvect}
\usepackage{color}
\usepackage{bm}
\usepackage{hyperref}

\newcommand{\eff}{\mathrm{eff}}
\renewcommand{\min}{\mathrm{min}}
\renewcommand{\max}{\mathrm{max}}

\newcommand{\true}{{\rm true}}
\newcommand{\Lap}{\mathrm{Lap}}

\newcommand{\rich}{\mathrm{rich}}
\newcommand{\poor}{\mathrm{poor}}

\newcommand{\bea}{\begin{eqnarray}}
\newcommand{\eea}{\end{eqnarray}}
\newcommand{\beas}{\begin{eqnarray*}}
\newcommand{\eeas}{\end{eqnarray*}}

\newcommand{\data}{{\rm data}}
\newcommand{\set}[1]{\left\{ #1 \right\}}

\newcommand{\braket}[1]{\left\langle #1 \right\rangle}

\definecolor{red}{rgb}{1,0,0}

\begin{document}


\title{Density estimation on small datasets} 


\author{Wei-Chia Chen, Ammar Tareen, Justin B. Kinney}
\email[Email correspondence to ]{jkinney@cshl.edu}
\affiliation{Simons Center for Quantitative Biology, Cold Spring Harbor Laboratory, Cold Spring Harbor, New York 11724, USA}


\begin{abstract}
How might a smooth probability distribution be estimated, with accurately quantified uncertainty, from a limited amount of sampled data? Here we describe a field-theoretic approach that addresses this problem remarkably well in one dimension, providing an exact nonparametric Bayesian posterior without relying on tunable parameters or large-data approximations. Strong non-Gaussian constraints, which require a non-perturbative treatment, are found to play a major role in reducing distribution uncertainty. A software implementation of this method is provided.  
\end{abstract}




\maketitle


The need to estimate smooth probability distributions from a limited number of samples is ubiquitous in data analysis \cite{Silverman:1986}. This ``density estimation'' problem also presents a fundamental conceptual challenge in statistical learning, important aspects of which remain unresolved. These outstanding problems are especially acute in the context of small datasets, where standard large-dataset approximations do not apply. Here we investigate the potential for Bayesian field theory, an area of statistical learning based on field-theoretic methods in physics \cite{Bialek:1996wr, Lemm:2003, Ensslin:2009, Ensslin:2013dy}, to estimate probability densities in this small data regime.   

Density estimation requires answering two distinct questions. First, what is the \textit{best} estimate for the underlying probability distribution? Second, what do other \textit{plausible} distributions look like? Ideally, one would like to answer these questions by first considering all possible distributions (regardless of mathematical form), then identifying those that fit the data while satisfying a transparent notion of smoothness. Such an approach should not require one to manually identify values for critical parameters, specify boundary conditions, or make invalid mathematical approximations in the small data regime. However, the most common density estimation approaches, including kernel density estimation (KDE) \cite{Silverman:1986} and Dirichlet process mixture modeling (DPMM)\cite{Muller:2015wj, Gelman:2013dr}, do not satisfy these requirements.

Previous work has described a Bayesian field theory approach, called Density Estimation using Field Theory (DEFT) \cite{Kinney:2014el, Kinney:2015vu}, for addressing the density estimation problem in low dimensions. DEFT satisfies all of the above criteria except for the last one: in \cite{Kinney:2014el, Kinney:2015vu}, an appeal to the large data regime was used to justify a Laplace approximation (i.e., a saddle-point approximation) of the Bayesian posterior. This approximation facilitated the sampling of an ensemble of plausible densities, as well as the identification of an optimal smoothness lengthscale. Independent but closely related work \cite{Riihimaki:2014du} has also relied heavily on this approximation. 

Here we investigate the performance of DEFT in the small data regime and find that the Laplace approximation advocated in prior work can be catastrophic. This is because non-Gaussian features of the DEFT posterior are critical for suppressing ``wisps'' -- large positive fluctuations that otherwise occur in posterior-sampled densities. We further find that these non-Gaussian effects cannot be addressed perturbatively using Feynman diagrams, as has been suggested in other Bayesian field theory contexts \cite{Ensslin:2009,Ensslin:2013dy}. These results are not specific to DEFT, but rather reflect the fundamentally nonperturbative nature of the density estimation problem.

Happily, we find that importance resampling \cite{Gelman:2013dr} can rapidly and effectively correct for the Laplace approximation. The resulting DEFT algorithm, which we have made available in robust and easy-to-use software, thus appears to satisfy all of the above requirements for an ideal density estimation method in one dimension. Tests of DEFT on simulated data show favorable performance relative to KDE and DPMM. We also illustrate the utility of DEFT on real data from the Large Hadron Collider \cite{CMSCollaboration:2012ta} and World Health Organization (WHO) \cite{WHO:2017ab} .


\begin{figure}[h!]
\includegraphics{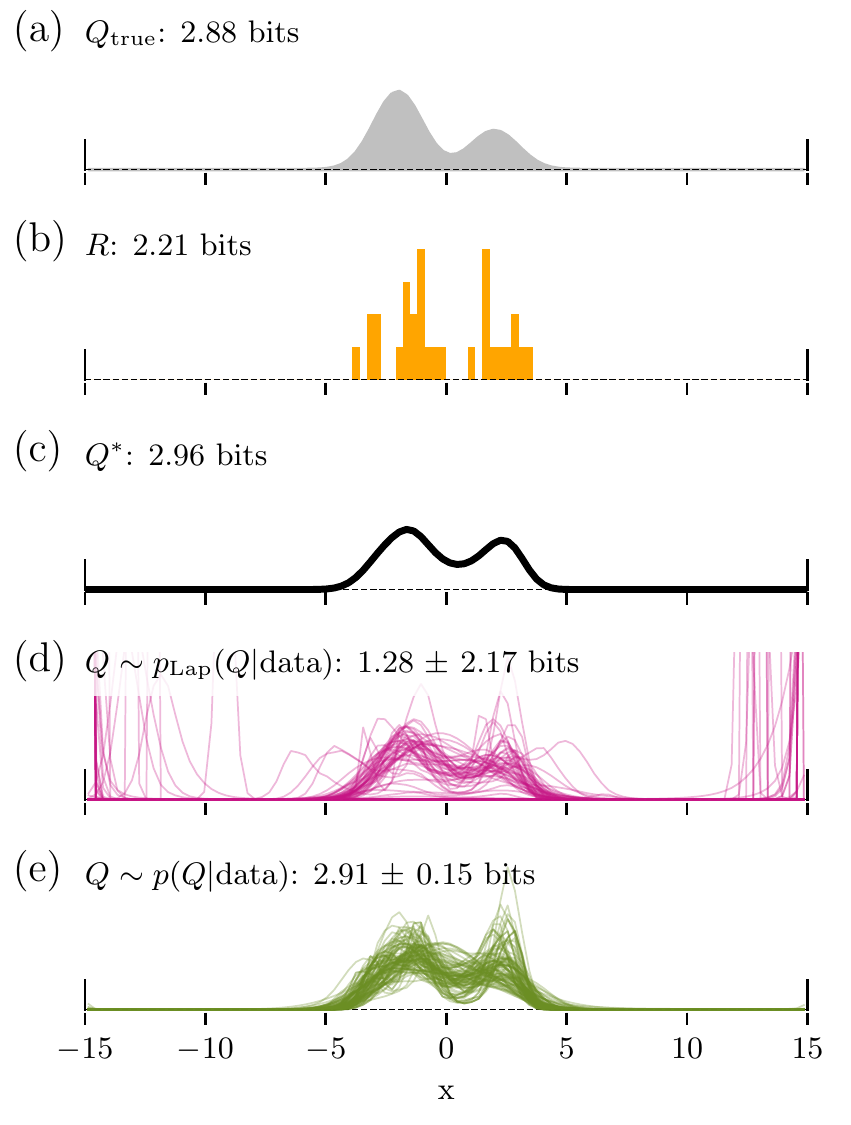}	
\caption{(Color) \textbf{Density  estimation using field theory}. (a) A Gaussian mixture distribution $Q_\true = \frac{2}{3} \mathcal{N}(-2,1) + \frac{1}{3} \mathcal{N}(2,1)$ within the $x$-interval $(-15,15)$.  (b) A histogram $R$ of $N=30$ data points sampled from $Q_\true$ and discretized to $G=100$ grid points. (c) The corresponding estimate $Q^*$ computed by DEFT using $\alpha = 3$ and the same grid as in (b). (d) $100$ distributions sampled from the Laplace-approximated posterior $p_\Lap(Q | \data)$, which accounts for uncertainty in $\ell$ as well as in $Q$. (e) $100$ distributions generated using importance resampling of the Laplace ensemble. The differential entropies of the illustrated distributions are provided. \label{fig:problem}} 
\end{figure}

We first recap the DEFT approach to density estimation \cite{Kinney:2014el, Kinney:2015vu}. Consider $N$ data points $\set{x_i}_{i=1}^N$ drawn from a smooth one-dimensional probability distribution $Q_\true(x)$ that is confined to an $x$-interval of length $L$. From these data we wish to obtain a best estimate $Q^*$ of $Q_\true$, as well as an ensemble of plausible distributions with which to quantify the uncertainty in this estimate.  

DEFT  reparametrizes each candidate distribution $Q$ in terms of a field $\phi$ via 
\begin{equation}
    Q(x) = \frac{e^{-\phi(x)}}{\int dx' e^{-\phi(x')}}.
\end{equation}
After adopting a Bayesian prior that constrains the $\alpha$-order $x$-derivative of $\phi$ (denoted by $\partial^\alpha \phi$ in what follows), and accounting for the likelihood of the data given $\phi$, one obtains a posterior distribution on $\phi$. We represent this posterior as $p(Q | \data, \ell) \propto \exp (-S_\ell[\phi])$ where
\begin{equation}
    S_\ell[\phi] =  \int \frac{dx}{L} \left[ \frac{\ell^{2 \alpha}}{2} (\partial^\alpha \phi)^2 + N R L \phi + N e^{-\phi}  \right] \label{eq:action}
\end{equation}
is the ``posterior action'' described in \cite{Kinney:2015vu}. In Eq.\ \ref{eq:action}, $\ell$ is a smoothness lengthscale that has yet to be determined and $R(x) = \frac{1}{N}\sum_{i=1}^N \delta(x-x_i)$ is a histogram  (of bin width zero) that summarizes the data. See Supplemental Information section SI.1 for details. The behavior of $Q$ under this action $S_\ell[\phi]$ is the primary focus of the present paper. 

$S_\ell[\phi]$ is minimized at the maximum a posteriori (MAP) field $\phi_\ell$. The MAP field $\phi_\ell$ is unique even in the absence of boundary conditions; see SI.2 for details. Although $\phi_\ell$  cannot be solved analytically, it is readily computed as the solution to a convex optimization problem after discretization of the $x$-domain at $G$ equally-spaced grid points. In this discrete representation, $R$ becomes a histogram with bin width $h = L/G$. As long as $h \ll \ell$, the choice of $G$ will not greatly affect $\phi_\ell$. The optimal lengthscale $\ell^*$ is identified by maximizing the Bayesian evidence, $p(\data | \ell)$; see SI.3 for details. $Q^* = Q_{\ell^*}$ is then used as our best density estimate. Fig.\ \ref{fig:problem}(a-c) illustrates this procedure on simulated data. 
 
To characterize the uncertainty in the DEFT estimate $Q^*$, we sample the Bayesian posterior $p(Q | \data) = \int d\ell\, p(\ell | \data) p(Q | \data, \ell)$. Each sample is generated by first drawing $\ell$ from $p(\ell | \data)$, then drawing $Q$ from $p(Q | \data, \ell)$. Previous work \cite{Kinney:2014el} has suggested that this sampling task be performed using the Laplace approximation, i.e., approximating $p(Q | \data, \ell)$ with a Gaussian that has the same mean and Hessian. The corresponding action, $S^\Lap_\ell[\phi]$, is thus quadratic in $\delta \phi = \phi - \phi_\ell$. This Laplace approximation has the advantage that posterior samples $Q$ can be rapidly and independently generated \cite{Kinney:2014el}. 

Fig.\ \ref{fig:problem}d shows multiple $Q$s sampled from the Laplace posterior $p_\Lap(Q | \data) = \int d\ell\, p(\ell | \data) p_\Lap(Q | \data, \ell)$. Clearly something is very wrong. Although many of these $Q$s appear reasonable, others exhibit wisps that have substantial probability mass far removed from the data.

We hypothesized that wisps are an artifact of the Laplace approximation. To correct for potential inaccuracies of this  approximation, we adopted an importance resampling approach \cite{Gelman:2013dr}. For each sampled $\phi$ we computed a weight
\begin{equation}
    w_\ell[\phi] = \exp \left( S^\Lap_\ell[\phi]-S_\ell[\phi] \right). \label{eq:weight}
\end{equation} 
We then resampled the Laplace ensemble with replacement, selecting each $\phi$ (and thus $Q$) with a probability proportional to $w_\ell[\phi]$. A mixture of such resampled ensembles across lengthscales $\ell$ was then used to generate an ensemble reflecting $p(Q | \data)$; see SI.4 for details. Fig.\ \ref{fig:problem}e shows 100 distributions $Q$ from this resampled posterior. Wisps no longer appear.

Eliminating wisps is especially important when estimating values for summary statistics, such as distribution entropy. In entropy estimation, the goal is to discern a value for the quantity $H_\true = H[Q_\true]$ where $H[Q] = - \int dx\, Q(x) \log_2 Q(x)$. Using the DEFT posterior ensemble, we can estimate $H_\true$ as $\widehat{H} \pm \widehat{\delta H}$, where $\widehat{H} = \braket{H}$ and $\widehat{\delta H} = \sqrt{\braket{H^2} - \braket{H}^2}$, with $\braket{\cdot}$ denoting a posterior average. Previous work expressed hope that the ensemble provided by the Laplace approximation might serve this purpose \cite{Kinney:2014el}. But in this case we see that $\widehat{H}$ is far less accurate than the point estimates $H[R]$ or $H[Q^*]$, and $\widehat{\delta H}$ is enormous (Fig.\ \ref{fig:problem}d). Importance resampling fixes both problems: the resulting $\widehat{H}$ is closer to $H_\true$ than either point estimate, and $\widehat{\delta H}$ is remarkably small (Fig. \ref{fig:problem}e).

We now turn to the problem of understanding how wisps arise. To this end we consider the variation in the action upon $\phi_\ell \to \phi_\ell + \delta \phi$. One finds that
\begin{equation}
    \delta S_\ell[\phi_\ell + \delta \phi] =  \int \frac{dx}{L} \frac{\ell^{2 \alpha}}{2} (\partial^\alpha \delta \phi)^2 + \int \frac{dx}{L} V(\delta \phi) \label{eq:two_parts}
\end{equation}
where
\begin{equation}
V(\delta \phi) = N L Q_\ell \left[ e^{-\delta \phi} - 1 + \delta \phi \right].  \label{eq:potential_exact}
\end{equation}
The first (kinetic) term on the right hand side of Eq.\ \ref{eq:two_parts} imposes a smoothness constraint on $\delta \phi$, while the second (potential) term keeps $\delta \phi$ confined to a potential well consistent with the data. See SI.5 for details. Note that $V$ is convex, nonnegative, and vanishes when $\delta \phi = 0$. By analogy to equipartition, we define  $n_\eff$, the effective number of degrees of freedom constrained by the data, as twice the value of the second term in Eq.\ \ref{eq:two_parts} averaged over the posterior ensemble. Typical fluctuations $\delta \phi$ will therefore exhibit $V(\delta \phi) \sim n_\eff/2$. 

We now separately consider  the ``data rich'' regime of the $x$ domain, which we define by $Q_\ell(x) \gg n_\eff/ 2NL$, and the ``data poor'' regime, corresponding to $Q_\ell(x) \ll n_\eff/ 2NL$. In the data rich regime,  fluctuations are small enough that $V$ adheres well to its Laplace approximation, $V \approx  N L Q_\ell \delta \phi^2 / 2$. Under this nearly symmetric potential, both positive fluctuations $\delta \phi^+$ and negative fluctuations $\delta \phi^-$ are constrained by
\begin{equation}
    |\delta \phi^{\pm}| \sim \delta \phi_\rich =  \sqrt{\frac{n_\eff}{N L Q_\ell}}. \label{eq:dphi_rich}
\end{equation}
By contrast, $V$ is highly asymmetric in the data poor regime and produces highly asymmetric fluctuations. Positive fluctuations satisfy $\delta \phi^+ \sim  n_\eff / 2NLQ_\ell$, whereas negative fluctuations obey
\begin{equation}
    - \delta \phi^- \sim \delta \phi^-_\poor = \log \frac{n_\eff}{2 N L Q_\ell}.  \label{eq:dphi_poor}
\end{equation}
See SI.5 for more information. 

\begin{figure}[t]
\includegraphics{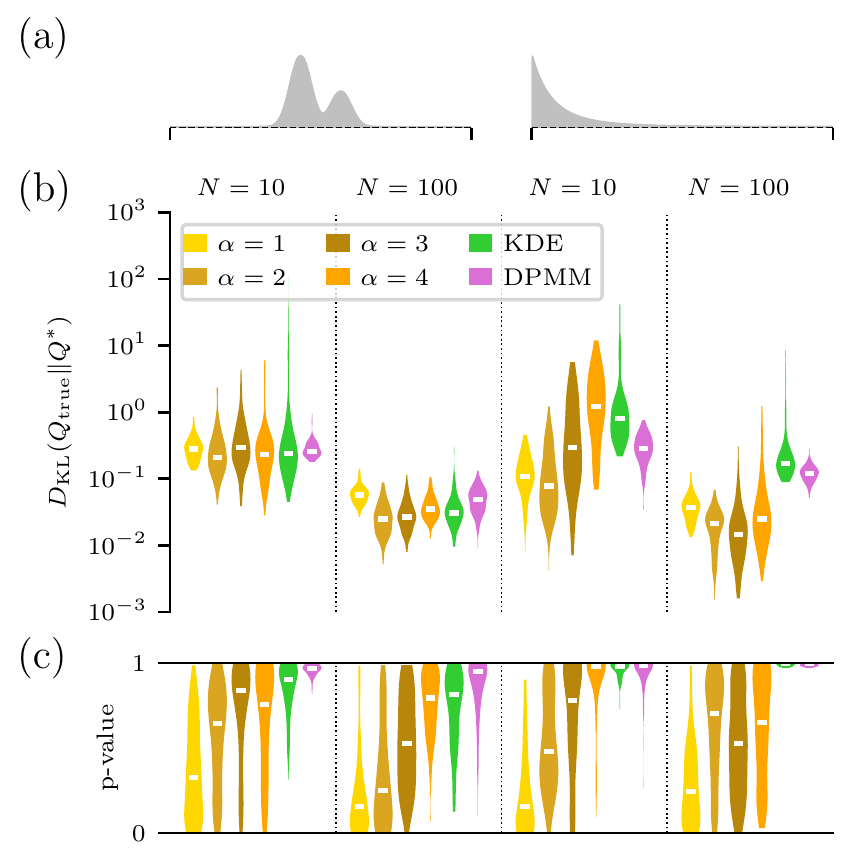}	
\caption{(Color) \textbf{Performance of DEFT}. (a) DEFT, KDE, and DPMM were used to analyze data from two different $Q_\true$ distributions: the Gaussian mixture from Fig.\ \ref{fig:problem}a (left) or a Pareto distribution, $Q_\true(x) = 3 x^{-4}$, confined to the $x$-interval $(1,4)$ (right). (b) 100 datasets of size $N=10$ and 100 datasets of size $N=100$ were generated for each $Q_\true$. For each dataset, $Q^*$ was computed by DEFT (using $G = 100$ and $\alpha =$ 1, 2, 3, or 4), by KDE, or by DPMM. Violin plots (with median indicated) show the resulting Kullback-Leibler divergences $D_\mathrm{KL}(Q_\true \| Q^*)$. (c) P-values quantifying, for each simulated dataset, the location of $D_\mathrm{KL}(Q_\true \| Q^*)$ within the distribution of $D_\mathrm{KL}(Q \| Q^*)$ values observed for $Q \sim p(Q | \data)$.  \label{fig:tests}} 
\end{figure}

The key point is that adopting $S^\Lap_\ell[\phi]$ in place of $S_\ell[\phi]$ is equivalent to assuming the Laplace approximation for $V$ throughout the entire $x$-domain. Because $\delta \phi_\rich \gg \delta \phi^-_\poor$ in data poor regions, the Laplace approximation greatly overestimates the size of downward fluctuations in $\phi$ . This results in the large upward fluctuations in $Q$ that we identify as wisps. We note that wisps are especially prominent at the $x$-interval boundaries in Fig.\ \ref{fig:problem} for two reasons: (i) $Q_\ell$ is especially small here, making these regions very data poor, and (ii) the kinetic term in Eq. \ref{eq:two_parts}, which is all that suppresses wisps in data poor regions, is less effective at constraining $\delta \phi$ because data are present on only one side.
 
 Feynman diagrams provide a general means of correcting for inaccuracies in Laplace approximations \cite{ZinnJustin:2010tt}, and have been advocated in the context of some Bayesian field theory regression problems \cite{Ensslin:2009, Ensslin:2013dy}. For density estimation, however, Feynman diagrams are ineffective if any region of the $x$ interval is data poor. This is due to the action $S_\ell[\phi]$ being strongly coupled.  For example, in the Bayesian evidence computations used to determine $\ell^*$, DEFT estimates the action $Z_\ell = \int \mathcal{D} \phi\, e^{-S_{\ell}[\phi]}$ using the Laplace approximation $Z_\ell^\Lap = \int \mathcal{D} \phi\, e^{-S^\Lap_{\ell}[\phi]}$. See SI.3 for details. At first, one might think it possible to correct for potential inaccuracies in this approximation using a series of vacuum diagrams (see SI.6), i.e.,
\begin{equation}
\log \frac{Z_\ell}{Z^\Lap_\ell} =
\begin{fmffile}{diagrams}
\fmfstraight
\parbox{5mm}{ 
\begin{fmfgraph*}(15,35) 
\fmfleft{l} \fmfright{r}
\fmf{phantom}{l,c,r} \fmffreeze
\fmf{plain,right,tension=0.4}{c,c}
\fmf{plain,left,tension=0.4}{c,c}
\fmfdot{c}
\end{fmfgraph*} 
} 
+~
\parbox{13mm}{ 
\begin{fmfgraph*}(35,35) 
\fmfleft{l} \fmfright{r}
\fmf{phantom}{l,a,b,r} \fmffreeze
\fmfdot{a,b}
\fmf{plain}{a,b} \fmffreeze
\fmf{plain,left}{l,a}
\fmf{plain,right}{l,a}
\fmf{plain,left}{b,r}
\fmf{plain,right}{b,r} 
\end{fmfgraph*} 
}
+
\parbox{10mm}{ 
\begin{fmfgraph*}(30,30) 
\fmfleft{l} \fmfright{r}
\fmf{phantom,tension=4}{l,a}
\fmf{phantom}{a,b}
\fmf{phantom,tension=4}{b,r} \fmffreeze
\fmf{plain}{a,b} 
\fmf{plain,left}{a,b}
\fmf{plain,right}{a,b}
\fmfdot{a,b} 
\end{fmfgraph*} 
}
\end{fmffile} 
+ \cdots. \label{eq:log_ratio_diagrams}
\end{equation}
However, as described in SI.8, the number of diagrams needed to obtain accurate results is prohibitive when data-poor regions of the $x$-interval are present. Fortunately, one can instead compute nonperturbative corrections to this log ratio using the importance resampling weights in Eq.\ \ref{eq:weight} via
\begin{equation}
    \log \frac{Z_\ell}{Z^\Lap_\ell} = \log \braket{w_\ell}_{\Lap|\ell}. \label{eq:log_ratio_sampling}
\end{equation}
See SI.7 for details. 
 
These results reflect a fundamental yet under-appreciated aspect of density estimation: unless data are observed throughout the  $x$-domain, the uncertainties in estimated probability densities require a nonperturbative treatment. Specifically, nonperturbative methods such as the Laplace approximation or Feynman diagrams can only be expected to work if $Q_\true(x) \gtrsim 1/NL$ everywhere within the $x$ domain.  Very often, however, density estimation is applied to data like that in Fig.\ \ref{fig:problem}, which is localized far away from one or both $x$-interval boundaries. We argue that the analysis of such data will quite generally require a nonperturbative treatment. 

To benchmark the performance of DEFT, we quantified its ability to estimate probability densities of known functional form. Specifically, we simulated datasets of varying size $N$ from a variety of $Q_\true$ distributions, then asked two questions. First, how accurately does $Q^*$ estimate $Q_\true$? Second, how typical is $Q_\true$ among the distributions in the Bayesian posterior? In both contexts, DEFT was compared to KDE and DPMM. See SI.9 for details on how KDE and DPMM were implemented. Fig.\ \ref{fig:tests} shows the results of these performance tests for two different choices of $Q_\true$. Fig.\ S3 in SI provides analogous results for other $Q_\true$ distributions. 

To answer the first question, we compared the Kullback-Leibler divergence,  $D_\mathrm{KL}(Q_\true \| Q^*)$, achieved by each estimator on each dataset. Note that smaller values for these divergences indicate better method accuracy. As illustrated in Fig.\ \ref{fig:tests}b, DEFT usually performed comparably to KDE and DPMM at $N$ = 10, and somewhat better at $N$ = 100. DEFT appears to have a particular advantage over both KDE and DPMM on $Q_\true$ distributions that bump up against one or both $x$-interval boundaries. Also unsurprising is that DEFT performs notably better with $\alpha$ = 2, 3, and 4 than with $\alpha$ = 1, since $\alpha$ = 1 yields non-smooth $Q^*$ distributions with cusps at each data point \cite{Kinney:2014el, Nemenman:2002}.

To answer the second question, we computed where $D_\mathrm{KL}(Q_\true \| Q^*)$ falls within the distribution of divergences $D_\mathrm{KL}(Q \| Q^*)$ observed for $Q \sim p(Q | \data)$. This location is naturally quantified by a p-value corresponding to the null hypothesis that $Q_\true \sim p(Q | \data)$. If $Q_\true$ is typical of plausible $Q$s, these p-values should be uniformly distributed between 0 and 1. Alternatively, p-values clustered close to 0 indicate that posterior ensemble $p(Q|\data)$ overestimates how much $Q_\true$ diverges from $Q^*$, whereas p-values clustered close to 1 indicate that $p(Q|\data)$ underestimates this uncertainty. Fig.\ \ref{fig:tests}c shows our results for the two choices of $Q_\true$ in Fig.\ \ref{fig:tests}a; results for other choices of $Q_\true$ are shown in Fig.\ S3. In general, the p-values for DEFT (with $\alpha$ = 2, 3, and 4) were distributed with remarkable uniformity. DEFT with $\alpha$ = 1 tended to overestimate uncertainties, whereas KDE and DPMM tended to underestimate uncertainties.


\begin{figure}[t]
\includegraphics{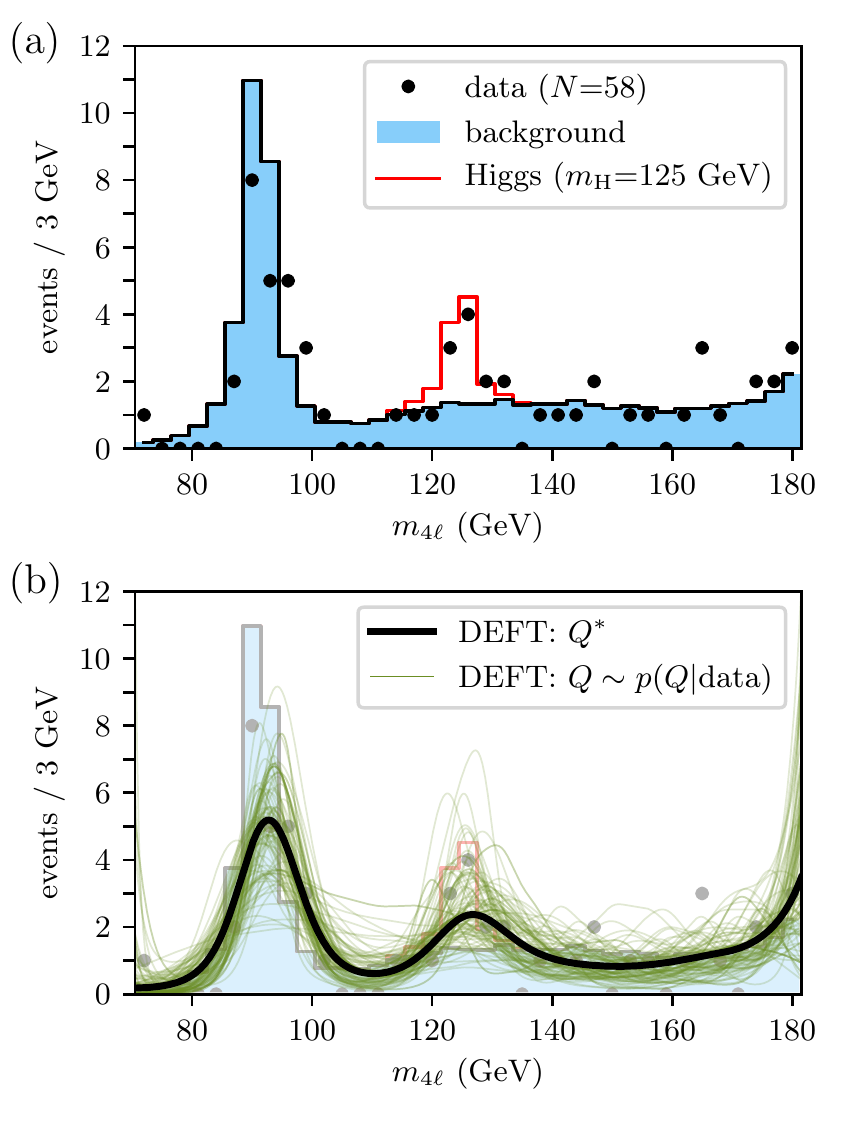}	
\caption{(Color) \textbf{DEFT applied to Higgs boson data.} (a) A reconstruction of Fig.\ 4 from \cite{CMSCollaboration:2012ta}. Dots (black) indicate the invariant masses of 4-lepton decay events histogrammed across $G = 37$ bins of width 3 GeV each. Also shown are the number of events expected, based on Standard Model simulations, from either background decay processes (blue) or from the decay of a Higgs boson with mass of 125 GeV (red). (b) The optimal density estimate $Q^*$ (black), along with 100 posterior samples $Q \sim p(Q | \data)$ (olive), computed by DEFT using the histogram data in panel (a).}
\label{fig:higgs}
\end{figure}

Finally, we illustrate the capabilities of DEFT using data reported in the initial observation of the Higgs boson \cite{CMSCollaboration:2012ta} (see Fig.\ S4 for an analysis of data from the WHO). Fig.\ \ref{fig:higgs}a, which is a reconstruction of  Fig.\ 4 of \cite{CMSCollaboration:2012ta}, shows a histogram of the invariant masses of $N=58$ 4-lepton events observed by the CMS Collaboration at the Large Hadron Collider. Such events are generated by the decays of the Higgs boson via H $\to$ ZZ $\to$ 4$\ell$, but they also arise from a variety of background decay processes.  One of the challenges faced by the CMS Collaboration was determining whether these data exhibit a localized excess of events representing a possible Higgs resonance. Fig.\ \ref{fig:higgs}b shows DEFT applied to these data using default parameters. Despite Higgs decays representing only $\sim 10\%$ of the observed events, DEFT detects a prominent local maxima near the Higgs resonance at $m_\mathrm{H}$ = 125 GeV. The confidence in this maxima can be quantified by sampling $p(Q|\data)$: 81\% of sampled $Q$s have exactly one local maximum between 110 GeV and 140 GeV (7\% have no local maxima and 12\% have multiple local maxima), and these maxima occurred at 127.1 GeV $\pm$ 3.7 GeV. 


Here we have shown that DEFT can effectively address density estimation needs on small datasets in one dimension. DEFT provides point estimates comparable to KDE and DPMM, but does not suffer from the multiple drawbacks of these other methods. In particular, the only key parameter that the user must specify is a small positive integer $\alpha$ that defines the qualitative meaning of smoothness and which governs how DEFT relates to maximum entropy estimation (see \cite{Kinney:2015vu}). In our experience, however, using $\alpha = 3$ seems to work well nearly all of the time.  Other parameters, such as the number of grid points $G$, reflect computational practicalities. These parameters can be chosen automatically and have little effect on the results as long as reasonable values are used.

DEFT thus addresses a major outstanding need, not just in statistical learning theory but also in day-to-day data analysis. To this end we have developed an open source Python package called SUFTware. SUFTware allows users to apply DEFT in one dimension to their own data, and in the future will include additional field-theory-based statistical methods. This implementation is sufficiently fast for routine use: the computations for Fig.\ \ref{fig:problem} takes about 0.25 seconds on a standard  laptop computer (see SI.10 for a discussion of computational complexity). SUFTware has minimal dependencies, is compatible with both Python 2 and Python 3, and is readily installed using the \texttt{pip} package manager. See \url{http://suftware.readthedocs.io} for installation and usage instructions.

We thank Kush Coshic for preliminary contributions to this project, as well as Serena Bradde, David McCandlish, and two anonymous referees for helpful feedback. This work was supported by a CSHL/Northwell Health Alliance grant to JBK and by NIH Cancer Center Support Grant 5P30CA045508.  

%
%
\bibliography{18_deft1d}

\cleardoublepage

%
%

\begin{widetext}

\renewcommand{\tocname}{Supplemental Information}

\setcounter{figure}{0}
\setcounter{equation}{0}
\renewcommand{\thefigure}{S\arabic{figure}}
\renewcommand\thesection{SI.\arabic{section}}
\renewcommand{\theequation}{S\arabic{equation}}


%
%
\section{The posterior action $S_\ell[\phi]$}

A derivation for Eq.\ 2 has already been reported in Ref.\ \cite{Kinney:2015vu}.  The derivation presented here, however, is more straight-forward. The action in Eq.\ 2 is given by
\begin{equation}
    S_\ell[\phi] = S^0_\ell[\phi] + S_\data[\phi], \label{eq:sum}
\end{equation}
where $S_\ell^0[\phi]$ is the ``prior action'', corresponding to a Bayesian prior $p(Q|\ell) \propto \exp(-S_\ell^0[\phi])$, while  $S_\data[\phi]$, the ``likelihood action'', is related to likelihood via $p(\data | Q) \propto \exp(-S_\data[\phi])$. DEFT uses a prior action of the form
\begin{equation}
S_\ell^0[\phi] = \int \frac{dx}{L} \frac{\ell^{2\alpha}}{2} (\partial^\alpha \phi)^2. \label{eq:prior_action}
\end{equation}
The parameter $\alpha$ reflects a fundamental choice in how one defines ``smoothness'', and $\ell$ is a lengthscale below which fluctuations in $\phi$ are strongly damped. The derivation of $S_\data[\phi]$ is as follows.  Suppose we are given $N$  data points drawn from a probability distribution $Q_\true(x)$ that is confined to the interval $[x_\min, x_\max]$. Label these data in order of increasing value as $x_1, x_2, \dots, x_N$. Next, imagine these data as being produced by a stochastic process in time, with $x$ being the time variable and $r(x)$ being the instantaneous emission rate. The likelihood of the data is then given by
\begin{eqnarray}
(dx)^N p(\data|r)  &=& 
    \left[e^{-\int_{x_\min}^{x_1} dx\,r(x)} \right] \cdot 
    \left[dx\,r(x_1)\right] \cdot 
    \left[e^{-\int_{x_1}^{x_2} dx\,r(x)} \right] \cdot
    \left[dx\,r(x_2) \right] \ \cdots \ 
    \left[dx\,r(x_N) \right] \cdot 
    \left[e^{-\int_{x_N}^{x_\max } dx\, r(x)} \right] \notag \\
    &=& (dx)^N \ \exp\left\{ -\int_{x_\min}^{x_\max} dx\, r(x) \right\} \ \prod_{i=1}^N r(x_i) \notag \\
    &=& (dx)^N \ \exp\left\{ -\int dx\, r(x) + \sum_{i=1}^N \log\,r(x_i)  \right\} \notag \\
    &=& (dx)^N \ \exp\left\{ -\int dx \left[ r(x) -  N R(x)\log\,r(x) \right] \right\}   \label{eq:likelihood_derivation}               
\end{eqnarray}
where $\int dx$ indicates integration over the entire $x$-domain and $R(x) = N^{-1} \sum_{i=1}^N \delta(x-x_i)$ is the raw data density referred to in the main text.  Next, we parametrize the emission rate $r(x)$ using the field $\phi(x)$ via
\begin{equation}
r(x) = \frac{N}{L} e^{-\phi(x)}. \label{eq:emission_rate}
\end{equation}
The probability density corresponding to this rate is
\begin{equation}
    Q(x) = \frac{r(x)}{\int dx'\,r(x')} = \frac{e^{-\phi(x)}}{\int dx'\,e^{-\phi(x')}},
\end{equation}
and so our definition of $\phi$ here is consistent with the definition of $\phi$ in the main text. We therefore see that the likelihood density in Eq.\ \ref{eq:likelihood_derivation} is given by $p(\text{data}|\phi) \propto \exp(-S_\data[\phi])$ where the corresponding action (after dropping the constant term $N \log (L/N)$) is,
\begin{equation}
S_\data[\phi] =  \int \frac{dx}{L}  \left[ N L R(x)\phi(x) + N e^{-\phi(x)} \right]. \label{eq:likelihood_action}
\end{equation}
Plugging Eq.\ \ref{eq:prior_action} and Eq.\ \ref{eq:likelihood_action} into Eq.\ \ref{eq:sum} gives Eq.\ 2 of the main text. Note the origin of the two terms in the integrand in Eq.\ \ref{eq:likelihood_action}: the term linear in $\phi$ comes from the exact locations of the $N$ data points, whereas the nonlinear term (which leads to such interesting behavior) comes from regions of the $x$ domain in which no data is observed. 

We briefly discuss a subtle issue with the above derivation. The probability distribution $Q(x)$ is invariant under additive shifts in the underlying field, i.e., $\phi(x) \to \phi(x) + c$ for any constant $c$. By contrast, the likelihood action $S_\data[\phi]$ is not invariant under such transformations. This difference is due Eq.\ \ref{eq:emission_rate} which, by specifying how the emission rate $r(x)$ relates to $\phi(x)$, introduces an additional assumption about how $\phi$ should be constrained by data. But although this additional assumption alters $p(\phi | \data)$, it does not alter $p(Q | \data)$. The more involved  derivation of $S_\ell[\phi]$ provided in Ref.\ \cite{Kinney:2015vu} demonstrates this fact explicitly. 

%
%
\section{The MAP field $\phi_\ell$}

To solve for $\phi_\ell$, the maximum a posteriori (MAP) field at lengthscale $\ell$, we set $\delta S_\ell / \delta \phi = 0$. The resulting equation of motion is
\begin{equation}
\ell^{2\alpha} \Delta^\alpha \phi_\ell + NLR - Ne^{-\phi_\ell} = 0. \label{eq:eom}
\end{equation}
The operator $\Delta^\alpha$ that appears here is the ``bilateral Laplacian",  which is described in Ref.\ \cite{Kinney:2015vu}. Briefly, $\Delta^\alpha$ is defined by the requirement that
\begin{equation}
\int dx \ \varphi \Delta^\alpha \phi = \int dx \ (\partial^\alpha \varphi) (\partial^\alpha \phi),
\end{equation}
for any two fields $\varphi$ and $\phi$.  This bilateral Laplacian is identical to the standard $\alpha$-order Laplacian $(-1)^\alpha \partial^{2\alpha}$ in the interior of the $x$-interval, but differs at the boundaries. Specifically, the standard $\alpha$-order Laplacian requires the additional specification of $\alpha$ boundary conditions in order to be self-adjoint. By contrast, the bilateral Laplacian is self-adjoint without the specification of any boundary conditions. The equation of motion, Eq.\ \ref{eq:eom}, thus has a unique solution without the need to assume any boundary conditions on $\phi$. See Ref.\ \cite{Kinney:2015vu} for more information. 

By integrating Eq.\ \ref{eq:eom} we find that $\int dx\ e^{-\phi_\ell(x)} = L$, due to $\int dx\ R(x) = 1$ and $\int dx \Delta^\alpha \phi_\ell = \int dx (\partial^\alpha 1)(\partial^\alpha \phi_\ell) = 0$. The MAP density $Q_\ell$ thus has a simple form:
\begin{equation}
    Q_\ell(x) = \frac{e^{-\phi_\ell(x)}}{L}.
\end{equation}
Similarly, multiplying Eq.\ \ref{eq:eom} on the left by $x^k$ for $k = 1, \ldots, \alpha-1$ and integrating reveals that
\begin{equation}
    \braket{x^k}_{Q_\ell} = \braket{x^k}_{R}, \label{eq:moments}
\end{equation}
i.e., the first $\alpha-1$ moments of $Q_\ell$ exactly match those of the data. 

As described in Ref.\ \cite{Kinney:2015vu}, DEFT computes the map field $\phi_\ell$ for a set of lengthscales ${\ell_0, \ell_1, \ell_2, \ldots, \ell_K}$, ranging from $\ell_0 = 0$ to $\ell_K = \infty$. These lengthscales are chosen so that neighboring MAP densities, $Q_{\ell_{k}}$ and $Q_{\ell_{k+1}}$, are approximately equally spaced along this ``MAP curve'', as quantified by the geodesic distance $D_\mathrm{geo} (Q_{\ell_k}, Q_{\ell_{k+1}})$. We note that $Q_0$ is in fact the data histogram $R$, while $Q_\infty$ is in fact the maximum entropy distribution consistent with the moment constraints in Eq.\ \ref{eq:moments}. See Ref.\ \cite{Kinney:2015vu} for details.

%
%
\section{The evidence \NoCaseChange{$p(\data|\ell)$}}

The DEFT algorithm computes the MAP field at lengthscales spanning $\ell = 0$ to $\ell = \infty$. The optimal lengthscale $\ell^*$ is then computed by maximizing the Bayesian evidence  $p(\text{data}|\ell)$. The key quantity needed for this procedure is the ``evidence ratio,''  which is given by
\begin{equation}
    E(\ell) = \frac{p(\data|\ell)}{p(\data|\infty)}.
\end{equation}
It can be shown that $E(\ell) = (Z_\ell/Z_\ell^0)/(Z_\infty/Z_\infty^0)$, where 
\begin{equation}
Z_\ell = \int \mathcal{D}\phi \ e^{-S_\ell[\phi]} \ \ \ \text{and} \ \ \ Z_\ell^0 = \int \mathcal{D}\phi \ e^{-S_\ell^0[\phi]}
\end{equation}
respectively denote the posterior partition function and the prior partition function. The prior partition function $Z_\ell^0 $ can be computed analytically, although it has a divergence that must be regularized. By contrast, the posterior partition function $Z_\ell$ can only be analytically computed in the Laplace approximation. We therefore instead use the quantity 
\begin{equation}
Z_\ell^{\text{Lap}} = \int \mathcal{D}\phi \ e^{-S_\ell^{\text{Lap}}[\phi]},
\end{equation}
where $S_\ell^\text{Lap}[\phi]$ is the Laplace approximation of $S_\ell[\phi]$. The resulting evidence ratio in this approximation is found to be
\begin{equation}
E(\ell) = e^{S_\infty[\phi_\infty] - S_\ell[\phi_\ell]}
          \sqrt{\frac{\text{det}_{\text{ker}}[e^{-\phi_\infty}] \text{det}_{\text{row}}[L^{2\alpha}\Delta^\alpha]}
                     {\eta^{-\alpha} \ \text{det}[L^{2\alpha}\Delta^\alpha + \eta e^{-\phi_\ell}]}},
\end{equation}
where $\eta = N(L/\ell)^{2\alpha}$, and ``ker'' and ``row'' respectively denote the kernel and row space of the bilateral Laplacian $\Delta^\alpha$. See Ref.\ \cite{Kinney:2015vu} for details.

It should be emphasized that, although the Laplace approximation can be grossly innacurate when sampling $Q \sim p(Q | \data)$, it does not strongly effect the evidence ratio $E(\ell)$. This is because $\log E(\ell)$ typically varies over many orders of magnitude, whereas $\log (Z_\ell/Z_\ell^\Lap)$ varies with $\ell$ far less dramatically. This is demonstrated in Fig.\ \ref{fig:perturbation} below. Nevertheless, the SUFTware implementation of DEFT includes an option to correct for this approximation using importance sampling, as described in the main text. 

%
%
\section{Sampling the posterior \NoCaseChange{$p(\phi, \ell | \data)$}}

The posterior probability $p(\phi,\ell|\text{data})$ can be decomposed as
\begin{equation}
p(\phi,\ell|\text{data}) = p(\phi|\ell,\text{data}) \ p(\ell|\text{data}).
\end{equation}
This forms the basis for our posterior sampling procedure. First, we sample plausible $\ell$s from $p(\ell|\text{data})$. Note that $p(\ell|\text{data}) \propto p(\text{data}|\ell) \ p(\ell)$ by Bayes's Theorem. Assuming $p(\ell)$ is uniform over the length of the MAP curve as quantified by geodesic distance (see Ref.\ \cite{Kinney:2015vu}), $p(\ell|\text{data})$ becomes proportional to the evidence ratio $E(\ell)$. We thus sample values of $\ell$ from the set $\set{\ell_0, \ell_1, \ldots, \ell_K}$ used to trace the MAP curve, each $\ell_k$ being selected with probability proportional to $E(\ell_k)$. For each of these $\ell$ values, we then sample plausible $\phi$s from $p(\phi|\ell,\text{data})$. Here we employ importance sampling. Specifically, we can rewrite the distribution $p(\phi|\ell,\text{data})$ as follows
\begin{equation}
p(\phi|\ell,\text{data}) = \frac{e^{-S_\ell[\phi]}}{Z_\ell}
                         = \frac{e^{-S_\ell^{\text{Lap}}[\phi]}}{Z_\ell^{\text{Lap}}}
                           \frac{w_\ell[\phi]}
                                {\langle w_\ell \rangle_{\Lap|\ell}}
                        \propto p_\Lap(\phi | \ell, \data) w_\ell[\phi],
\end{equation}
where we have made use of Eq.\ \ref{eq:thing} (derived below). Therefore, we first sample $\phi$s from the Laplace-approximated distribution $p_\Lap(\phi | \data, \ell)$, then correct for the non-Gaussian nature of the original distribution by resampling these $\phi$s using the importance weights $w_\ell[\phi]$. 

%
%
\section{Origin of wisps}

\begin{figure}[t]
\centering
\includegraphics{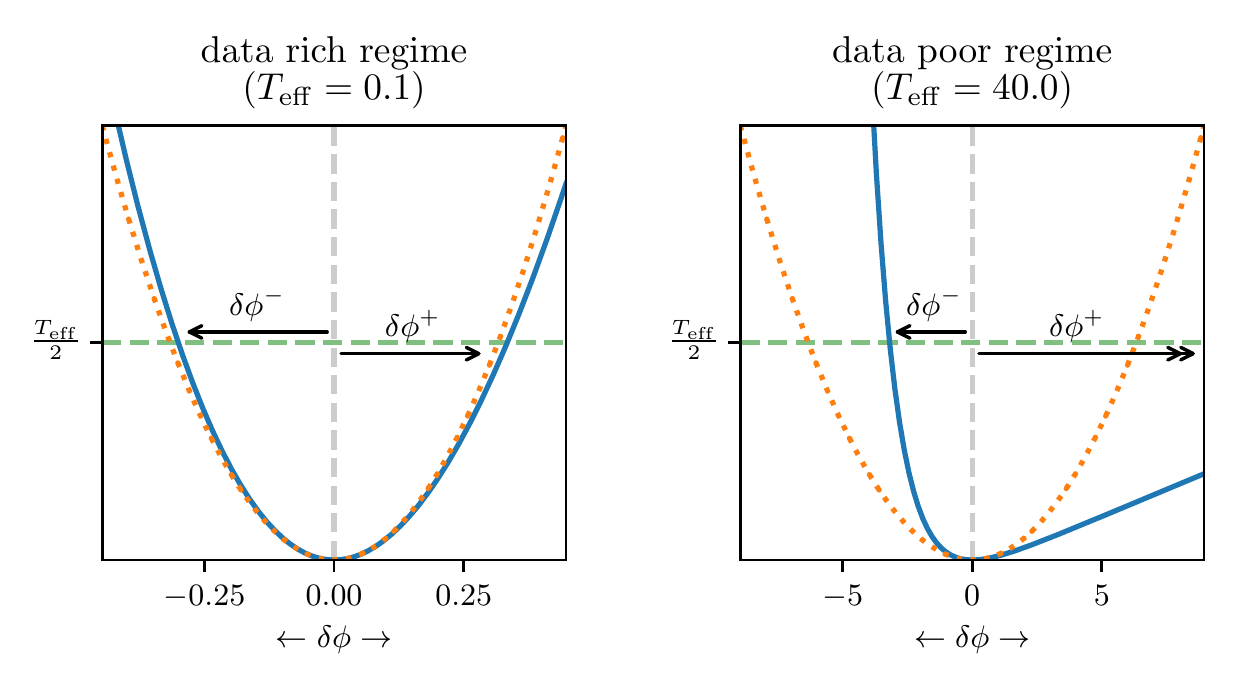}	
\caption{(Color) \textbf{Fluctuations $\delta \phi$ in data rich vs.\ data poor regimes}. Solid blue lines indicate $f(\delta \phi)$ from Eq.\ \ref{eq:f}. Dotted orange lines indicate the Laplace approximation $f_\Lap(\delta \phi) = \delta \phi^2/2$. Dashed green lines indicate the value of half the effective temperature ($T_\eff / 2$) from Eq.\ \ref{eq:beta}. In the data rich regime, $T_\eff/2 \ll 1$ (left panel), resulting in nearly symmetric $\delta \phi^{\pm}$ fluctuations. In the data poor regime, $T_\eff/2 \gg 1$ (right panel), resulting in highly asymmetric fluctuations; in particular, the magnitude of $\delta \phi^-$ due to $f$ is substantially less than would result from $f_\Lap$. Note also that the very large positive fluctuations $\delta \phi^+$ in the data poor regime have little noticeable effect on $Q_\ell$, since they just push $Q_\ell$ closer to zero.
\label{fig:fluctuations}} 
\end{figure}

To derive Eqs.\ 4 and 5, it suffices to note that 
\begin{eqnarray}
    S_\ell[\phi_\ell + \delta \phi] &=& S_\ell[\phi_\ell] + O(\delta \phi^2),
\end{eqnarray}
because the EOM in Eq.\ \ref{eq:eom} causes all first-order terms in $\delta \phi$ to cancel. Next, we express $V(\delta \phi) = N L Q_\ell f(\delta \phi)$ where
\begin{equation}
    f(\delta \phi) = e^{-\delta \phi} - 1 + \delta \phi \label{eq:f}
\end{equation}
is just $e^{-\delta \phi}$ with the 0th and 1st order terms subtracted out. This function is plotted in Fig.\ \ref{fig:fluctuations}. The key to deriving the magnitude of fluctuations $\delta \phi$ in different regimes is the relationship $\braket{V(\delta \phi)} \sim \frac{n_\eff}{2}$, which we rephrase here as
\begin{equation}
    \braket{f(\delta \phi)} \sim  \frac{T_\eff}{2}. 
\end{equation}
where 
\begin{equation}
    T_\eff = \frac{n_\eff}{N L Q_\ell} \label{eq:beta}
\end{equation} 
is an effective temperature. 

In the data rich regime, $T_\eff/2 \ll 1$. Therefore, $f(\delta \phi) \ll 1$ for typical fluctuations $\delta \phi$. As illustrated in Fig.\ \ref{fig:fluctuations} (left panel), the Laplace approximation  works well in this regime. Setting
\begin{equation}
    f(\delta \phi) \approx f_\Lap(\delta \phi) = \frac{\delta \phi^2}{2} \sim \frac{T_\eff}{2}
\end{equation}
and solving for $\delta \phi$ gives Eq.\ 6. 

In the data poor regime, $T_\eff/2 \gg 1$. As illustrated in Fig.\ \ref{fig:fluctuations} (right panel), $f$ is highly asymmetric in this regime and so the positive and negative fluctuations, $\delta \phi^+$ and $\delta \phi^-$, need to be treated separately. Specifically, 
\begin{equation}
    f(\delta \phi^+) \approx \delta \phi^+ \sim \frac{T_\eff}{2},~~~ \mathrm{whereas}~~~
    f(\delta \phi^-) \approx e^{-\delta \phi^-} \sim \frac{T_\eff}{2}.
\end{equation}
Solving the latter condition for $\delta \phi^-$ gives Eq.\ 7. Note in Fig.\ \ref{fig:fluctuations} (right panel) how the the Laplace approximation greatly overestimates the magnitude of negative fluctuations $\delta \phi^-$ in the data poor regime.

%
%
\section{Computing \NoCaseChange{$\log (Z_\ell/Z_\ell^\Lap)$} using Feynman diagrams}

Here we show how Feynman diagrams can be used to compute $\log (Z_\ell/Z_\ell^\Lap)$, thereby obtaining corrections to the Laplace approximation. Our exposition closely follows that sketched by Zinn-Justin \cite{ZinnJustin:2010tt}. However, because Feynman diagrams are rarely used in the context of statistical inference, we felt it worthwhile to make these calculations explicit. 

Upon discretization of the $x$-interval using $G$ grid points, the action in Eq.\ 2 becomes
\begin{equation}
S_\ell[\phi] = \frac{\ell^{2\alpha}}{2G} \sum_{ij} \Delta_{ij}^\alpha \phi_i \phi_j
             + \frac{NL}{G} \sum_i R_i \phi_i 
             + \frac{N}{G} \sum_i e^{-\phi_i}. 
\end{equation}
where $i, j = 1, 2, \ldots, G$. In what follows we represent fluctuations in $\phi$ about from the MAP field $\phi^\ell$ using the rescaled fluctuation $x = \sqrt{N} (\phi - \phi^\ell)$. The action can then be expanded in the following way:
\begin{equation}
S_\ell[\phi] = S_\ell^\text{Lap}[\phi] 
             + \frac{1}{3!} \sum_{ijk} \frac{B_{ijk}}{\sqrt{N}} x_i x_j x_k 
             + \frac{1}{4!} \sum_{ijkl} \frac{C_{ijkl}}{N} x_i x_j x_k x_l + \cdots, \label{eq:expansion}
\end{equation}
where the Laplace action is
\begin{equation}
S_\ell^\text{Lap}[\phi] = S_\ell[\phi^\ell] + \frac{1}{2} \sum_{ij} A_{ij} x_i x_j,
\end{equation}
and
\begin{eqnarray}
A_{ij} = \frac{1}{N} \left. \frac{\partial^2 S_\ell}{\partial\phi_i \partial\phi_j} \right|_{\phi^\ell}
&=& \frac{\ell^{2\alpha}}{NG} \Delta_{ij}^{\alpha} + \frac{1}{G} e^{-\phi_i^\ell} \delta_{ij}, \\
B_{ijk} = \frac{1}{N} \left. \frac{\partial^3 S_\ell}{\partial\phi_i \partial\phi_j \partial\phi_k} \right|_{\phi^\ell}
&=& - \frac{1}{G} e^{-\phi_i^\ell} \delta_{ijk}, \\
C_{ijkl} = \frac{1}{N} \left. \frac{\partial^4 S_\ell}{\partial\phi_i \partial\phi_j \partial\phi_k \partial\phi_l} \right|_{\phi^\ell}
&=& \frac{1}{G} e^{-\phi_i^\ell} \delta_{ijkl}.
\end{eqnarray}
The quantity $\log(Z_\ell/Z_\ell^\text{Lap})$ is conveniently given by the sum of connected vacuum diagrams. At $O(N^{-1})$, the relevant diagrams contain only 3rd-order and 4th-order vertices. From the expansion in Eq.\ \ref{eq:expansion} we see that the values corresponding to these vertices are given by $-B_{ijk}/\sqrt{N}$ and $-C_{ijkl}/N$, respectively. We also need the propagator matrix $P$, which is given by the inverse of the Hessian $A$, i.e., $P_{ij} = (A^{-1})_{ij}$. We thus obtain
\begin{equation}
\log \frac{Z_\ell}{Z^{\text{Lap}}_\ell} =
\begin{fmffile}{diagrams}
\fmfstraight
\parbox{5mm}{ 
\begin{fmfgraph*}(15,35)\fmfkeep{diagram1}
\fmfleft{l} \fmfright{r}
\fmf{phantom}{l,c,r} \fmffreeze
\fmf{plain,right,tension=0.4}{c,c}
\fmf{plain,left,tension=0.4}{c,c}
\fmfdot{c}
\end{fmfgraph*} 
} 
+~
\parbox{13mm}{ 
\begin{fmfgraph*}(35,35)\fmfkeep{diagram2} 
\fmfleft{l} \fmfright{r}
\fmf{phantom}{l,a,b,r} \fmffreeze
\fmfdot{a,b}
\fmf{plain}{a,b} \fmffreeze
\fmf{plain,left}{l,a}
\fmf{plain,right}{l,a}
\fmf{plain,left}{b,r}
\fmf{plain,right}{b,r} 
\end{fmfgraph*} 
}
+
\parbox{10mm}{ 
\begin{fmfgraph*}(30,30)\fmfkeep{diagram3} 
\fmfleft{l} \fmfright{r}
\fmf{phantom,tension=4}{l,a}
\fmf{phantom}{a,b}
\fmf{phantom,tension=4}{b,r} \fmffreeze
\fmf{plain}{a,b} 
\fmf{plain,left}{a,b}
\fmf{plain,right}{a,b}
\fmfdot{a,b}
\end{fmfgraph*} 
}
\end{fmffile}
+ O(N^{-2}),
\end{equation}
where the contribution from each diagram is
\begin{eqnarray}
\parbox{5mm}{\fmfreuse{diagram1}}
&=& \frac{1}{8} \sum_{ijkl} \left( -\frac{C_{ijkl}}{N} \right) P_{ij} P_{kl} 
= -\sum_{i} \frac{e^{-\phi^\ell_{i}}}{8NG} \left( P_{ii} \right)^2, \\
\parbox{13mm}{\fmfreuse{diagram2}}
&=& \frac{1}{8} \sum_{ijk} \sum_{lmn} \left( -\frac{B_{ijk}}{\sqrt{N}} \right) \left( -\frac{B_{lmn}}{\sqrt{N}} \right)
                                          P_{ij} P_{kl} P_{mn}
= \sum_{i} \sum_{l} \frac{e^{-\phi_{i}^\ell-\phi_{l}^\ell}}{8NG^2} P_{ii} P_{il} P_{ll}, \\
\parbox{10mm}{\fmfreuse{diagram3}}
&=& \frac{1}{12} \sum_{ijk} \sum_{lmn} \left( -\frac{B_{ijk}}{\sqrt{N}} \right) \left( -\frac{B_{lmn}}{\sqrt{N}} \right)
                                          P_{il} P_{jm} P_{kn}
= \sum_{i} \sum_{l} \frac{e^{-\phi_{i}^\ell-\phi_{l}^\ell}}{12NG^2} \left( P_{il} \right)^3.
\end{eqnarray}

%
%
\section{Computing \NoCaseChange{$\log (Z_\ell/Z_\ell^\Lap)$} using importance sampling}

Alternatively, the correction $\log (Z_\ell/Z_\ell^\Lap)$ can be computed using importance sampling involving the weights $w_\ell$ in Eq.\ 3. To see how, we express the partition function $Z_\ell$ as an average over the Laplace ensemble:
\begin{eqnarray}
Z_\ell &=& \int \mathcal{D}\phi \ e^{-S_\ell[\phi]} \\
       &=& Z_\ell^\text{Lap} \int \mathcal{D}\phi \ \frac{e^{-S_\ell^{\text{Lap}}[\phi]}}{Z_\ell^\text{Lap}} \ e^{S_\ell^{\text{Lap}}[\phi] - S_\ell[\phi]} \\
       &=& Z_\ell^\text{Lap} \int \mathcal{D}\phi \ p_\Lap(\phi | \data, \ell) \ w_\ell[\phi]\\
       &=& Z_\ell^{\text{Lap}} \ \langle w_\ell \rangle_{\text{Lap}|\ell}, \label{eq:thing}
\end{eqnarray}
where $\langle \cdot \rangle_{\text{Lap}|\ell}$ denotes the mean taken with respect to the Laplace posterior $p_\Lap(\phi | \data, \ell)$, and $w_\ell$ denotes the importance sampling weights in Eq.\ 3. The quantity $\log(Z_\ell/Z_\ell^\text{Lap})$ can thus be computed using Eq.\ 9.

\section{Feynman diagrams vs.\ importance sampling}

\begin{figure}[t]
\centering
\includegraphics{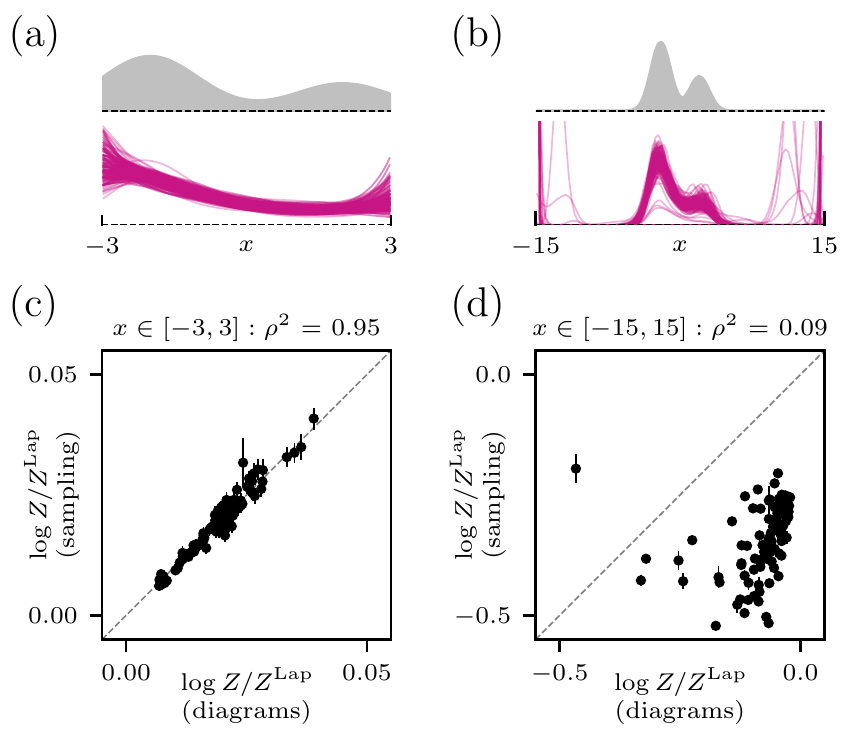}	
\caption{(Color) \textbf{Wisps appear when $S_\ell[\phi]$ is strongly coupled}. The accuracy of Feynman diagrams was assessed using data drawn from the $Q_\true$ density in Fig.\ 1 confined to the intervals $[-3,3]$ (a,c) or $[-15,15]$ (b,d). (a,b) $Q_\true$ (gray) is shown along with 100 distributions $Q$ (magenta) sampled at fixed $\ell = \ell^*$ from the Laplace-approximated posterior inferred from a dataset of size $N=100$. Wisps are observed in (b) but not in (a). (c,d) Values for $\log (Z_\ell/Z^\Lap_\ell)$ computed for 100 different datasets, generated as above, using either Feynman diagrams (Eq.\ 8) or importance weights (Eq.\ 9). These two quantities agree well in (c) but poorly in (d). Squared Pearson correlations, $\rho^2$, are shown in the titles of (c,d). 
\label{fig:perturbation}} 
\end{figure}

Perhaps disappointingly, Feynman diagrams generally do not work well in situations where wisps appear. This is because the posterior action in such cases is strongly coupled. To see this, consider an expansion of the potential $V$ in Eq.\ 5 to $m$'th order in $\delta \phi$:
\begin{equation}
    V_m(\delta \phi) = N L Q_\ell \sum_{n=2}^m \frac{(-\delta \phi)^n}{n!}. \label{eq:Vm}
\end{equation}
To produce accurate results, the potential $V_m$ must include enough terms to sufficiently approximate $V$ when evaluated at $\delta \phi = -\delta \phi^-_\poor = -\phi^* + \log (N/n_\eff)$. This would require $m_\mathrm{min} = \phi^* - \log(N/n_\eff)$ terms at the very least, since not until here do the (all positive) terms in this power series begin to decrease. Thus, the number of terms that would be needed cannot be fixed \emph{a priori}, but rather must increase with $\phi^*$. This presents a major problem for Feynman-diagram-based expansions.  Any diagram influenced by the the $m_\min$'th term in Eq.\ \ref{eq:Vm} must contain an $m_\min$'th order vertex. But $m_\min$ can be quite large: for $\phi^*$ in Fig.\ 1, finds $m_\min > 100$ near the boundaries of the $x$-interval. Evaluating Feynman diagrams up to such high order is not feasible.

This expectation is confirmed in Fig.\ \ref{fig:perturbation}, which compares the two ways of computing $\log(Z_\ell/Z_\ell^\text{Lap})$ for two different choices of $Q_\text{true}$. The Feynman diagram approximation works well when $Q_\text{true}$ fills the entire $x$-interval, indicating that the action $S_\ell[\phi]$ is nearly quadratic and the corrections to Laplace approximation are small. However, when $Q_\text{true}$ vanishes in large regions of the $x$ domain, the Feynman diagram approximation is a very bad. In this case, the action $S_\ell[\phi]$ is strongly coupled and a fundamentally non-perturbative approach is required to compute the corrections. 

Although the non-quadratic nature of the posterior action can lead to a partition function $Z_\ell$ differing from its Laplace-approximated value $Z_\ell^\text{Lap}$ by a large amount, we find that Laplace approximation generally works well nevertheless for identifying the optimal lengthscale. This is because $\log Z_\ell^\text{Lap}$ typically varies by multiple orders of magnitude across different values of $\ell$, thereby swamping potential inaccuracies in the $Z_\ell \approx Z_\ell^\text{Lap}$ assumption. 

%
%
\section{Other density estimation methods}

Here we describe the Kernel density estimation (KDE) and Dirichlet process mixture modeling (DPMM) algorithms used for the computations shown in Fig.\ 2 and Fig.\ \ref{fig:distributions}. 

\subsection{Kernel density estimation}

KDE is arguably the most common approach to density estimation in one dimension. Given data $\{x_i\}_{i=1}^N$, the KDE density estimate is given by
\begin{equation}
Q^*(x) = \frac{1}{N} \sum_{i=1}^{N} \frac{1}{w} K \left( \frac{x-x_i}{w} \right), \label{eq:kde}
\end{equation}  
where $K(z)$ is the kernel function and $w$ is the ``bandwidth''. We used a Gaussian kernel,
\begin{equation}
    K(z) = \frac{1}{\sqrt{2 \pi}} e^{-z^2/2},
\end{equation}
and chose the bandwidth $w$ using cross-validation. Specifically, we considered 100 candidate bandwidths geometrically distributed between $w_\min$ (the minimum spacing between data points) and $w_\max$ (10 times the span of the data). We then chose the bandwidth $w$ that maximized the jackknifed log likelihood
\begin{equation}
    \mathcal{L}(w) = \sum_{i=1}^N \log Q_{-i}^*(x_i),
\end{equation}
where the subscript on $Q^*_{-i}$ indicates the density $Q^*$ computed as Eq.\ \ref{eq:kde} but using a dataset missing the datum $x_i$. 

KDE does not provide an explicit posterior on $Q$. Therefore, to compute p-values for Fig.\ 2 and Fig.\ \ref{fig:distributions}, we approximated posterior samples $Q \sim p(Q | \data)$ by applying KDE to bootstrap-resampled datasets. 

\subsection{Dirichlet process mixture modeling}

DPMM is arguably the most popular nonparametric Bayesian method for estimating probability densities. DPMMs have a hierarchical structure, in the sense that each data point is assumed to be drawn from one of a number of ``clusters,'' with each cluster having a  probability density defined by a kernel of pre-specified functional form. 

In the computations for Fig.\ 2 and Fig.\ \ref{fig:distributions}, we adopted the finite DPMM described in Refs.\ \cite{Muller:2015wj,Gelman:2013dr}. Densities were assumed to be of the form
\begin{equation}
Q(x) = \sum_{h=1}^{H} w_h K_{m_h}(x),
\end{equation}
where $H$ is the number of clusters, $w_h$ is the probability of cluster $h$, and $m_h$ is the set of parameters defining the density of cluster $h$.  $K_m(z)$ was assumed to be a Gaussian density specified by $m = (\mu, \sigma^2)$, i.e., a mean and a variance.  A normal-inverse-gamma distribution was used as the prior on $m$: 
\begin{equation}
		p(\mu, \sigma^2) = \mathcal{N}(\mu|\hat{\mu},\hat{\kappa}\sigma^2) \ \Gamma^{-1}(\sigma^2|\hat{\alpha},\hat{\beta}),\label{eq:dpmmprior}
\end{equation}
where $\hat{\kappa} = 1$, $\hat{\alpha} = 1$,  $\hat{\beta} = \hat{\sigma}^2$,
\begin{equation}
    \hat{\sigma}^2 = \frac{1}{N-1} \sum_{i=1}^N (x_i - \hat{\mu})^2,~~\mathrm{and}~~\hat{\mu} = \frac{1}{N}\sum_{i=1}^N x_i. \label{eq:mean_and_variance}
\end{equation}
The number of clusters was fixed at $H = 10$. For each dataset, we used Gibbs sampling to obtain an ensemble of plausible densities representing $p(Q|\data)$. The optimal estimate $Q^*$ was then defined as the mean density in this ensemble. Following Ref.\ \cite{Gelman:2013dr}, our Gibbs sampling algorithm worked as follows. For each cluster $h = 1, 2, \ldots, H$, we chose an initial weight $w_h = 1/H$ and a set of kernel parameters $m_h$ chosen according to the prior distribution $p(\mu, \sigma^2)$ in Eq.\ \ref{eq:dpmmprior}. The sampler was then run by iterating the following steps:

\begin{enumerate}
\item Data were redistributed across clusters. Specifically, each data point $x_i$ was allocated to cluster $h$ with probability 
	\begin{equation}
		p(h | x_i) = \frac{w_h K_{m_h}(x_i)}{\sum_{h'=1}^H w_{h'} K_{m_{h'}}(x_i)}.	
	\end{equation}
\item The mean and variance of each cluster were updated using
	\begin{equation}
		m_h \sim \mathcal{N}(\mu_h|\hat{\mu}_h,\hat{\kappa}_h \sigma_h^2) \ \Gamma^{-1}(\sigma_h^2|\hat{\alpha}_h,\hat{\beta}_h),
	\end{equation}  
	where
	\begin{eqnarray}
		\hat{\mu}_h &=& \hat{\kappa}_h\left( \frac{\hat{\mu}}{\hat{\kappa}} + n_h \langle x_h \rangle \right), \\
		\hat{\kappa}_h &=& \frac{\hat{\kappa}}{1 + n_h \hat{\kappa}}, \\
		\hat{\alpha}_h &=& \hat{\alpha} + \frac{n_h}{2}, \\
		\hat{\beta}_h &=& \hat{\beta} + \frac{1}{2} 
		\left( \sum_{i \in h} (x_i - \langle x_h \rangle)^2 
		+ \frac{n_h}{1 + n_h \hat{\kappa}} (\langle x_h \rangle - \hat{\mu})^2 \right).
	\end{eqnarray} 
	Here, $x_h$ represents the set of data points belonging to cluster $h$ and $n_h = |x_h|$.
\item The cluster weights were updated by sampling 
	\begin{equation}
		w_1, \ldots, w_H \sim \mathrm{Dirichlet}(1+n_1, \ldots, 1+n_H).
	\end{equation}

\end{enumerate}

%
%
\section{\ \ Computational complexity}

An explicit expression for the algorithmic complexity of DEFT is not very helpful for understanding runtime performance. This is because DEFT involves multiple steps computed in series, the runtimes of which are governed by different parameters. In practice, we have found DEFT to be primarily limited by the number of grid points $G$. This is because a computation of the evidence ratio $E(\ell)$, as well as posterior sampling, requires a spectral decomposition of the $G \times G$ Hessian matrix at each lengthscale $\ell$ along the MAP curve. We note, however, that DEFT computations with $G = 100$ are generally quite fast (i.e., $\sim 0.25$ seconds on a standard laptop computer). Although DEFT does require histogramming the data, which is $\mathcal{O}(N)$, this is rarely the bottleneck in practice.  In fact, we have found that the speed of DEFT often \emph{increases} with $N$, since this leads to a shorter MAP curve, thus requiring fewer discrete lengthscales $\ell$ to be examined.

In our computations for Fig.\ 2 and Fig.\ \ref{fig:distributions}, DEFT was often faster than our KDE and DPMM implementations. The use of jackknife cross-validation greatly slows down KDE in a manner that increases linearly with $N$. DPMM, on the other hand, is greatly slowed down by its reliance on Gibbs sampling, which is necessitated by the non-convexity of the parameter posterior. In fact, Gibbs sampling is needed not just to generate a posterior sample, but also to estimate $Q^*$ (via a posterior mean). We note that the accuracy of KDE and DPMM is also very sensitive to the choice of kernel, especially when data is clustered near the $x$-interval boundaries. 

\clearpage
\begin{figure}[t]
\centering
\includegraphics[height=8in]{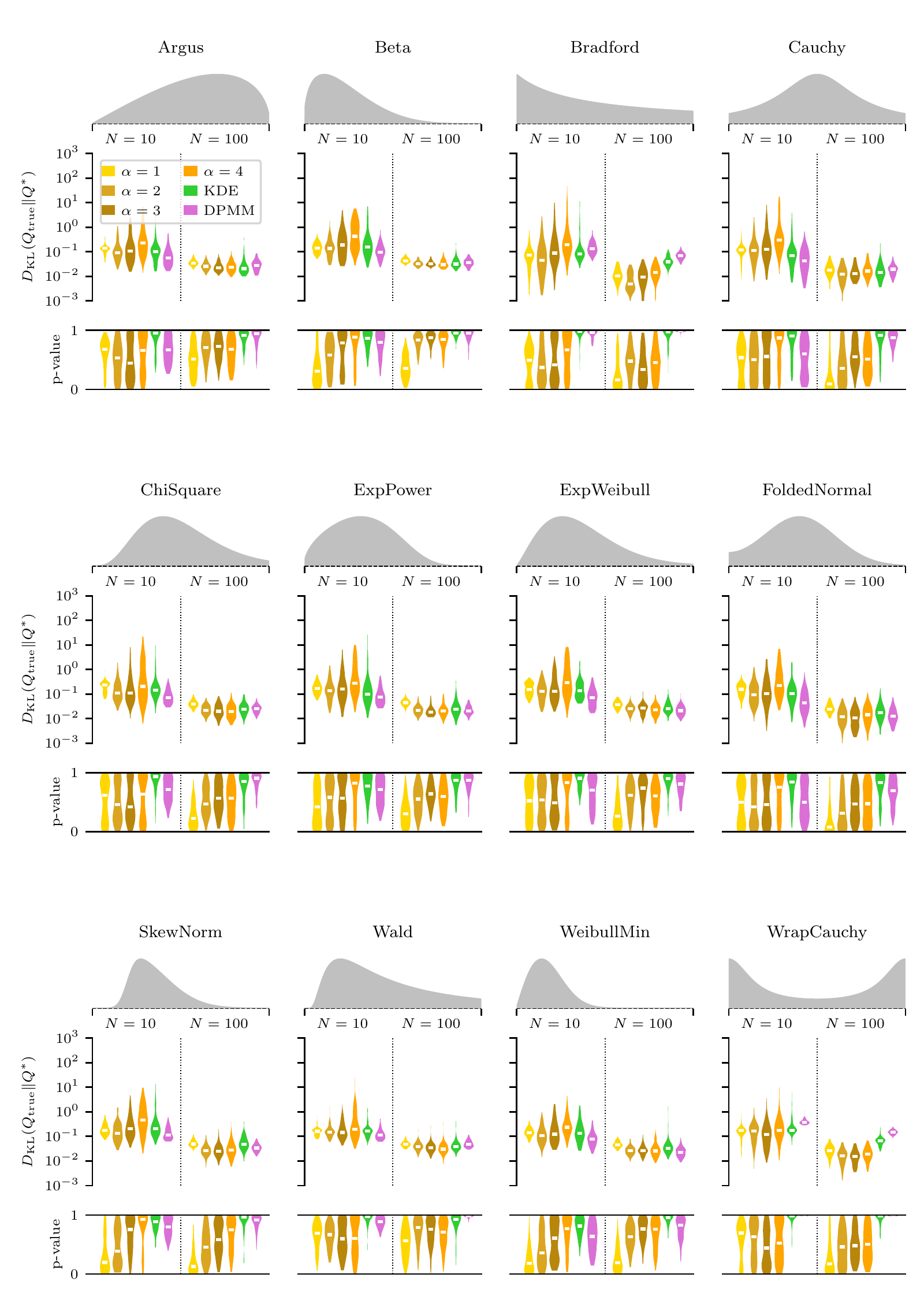}	
\caption{(Color) \textbf{Extension of Fig.\ 2 to other choices of $Q_\true$}. The same analysis as in Fig.\ 2 was performed for twelve additional $Q_\true$ distributions, which were selected from the built-in distributions  in the \texttt{scipy.stats} Python library.
\label{fig:distributions}} 
\end{figure}
    
\clearpage    
\begin{figure}[t]
\centering
\includegraphics[height=7in]{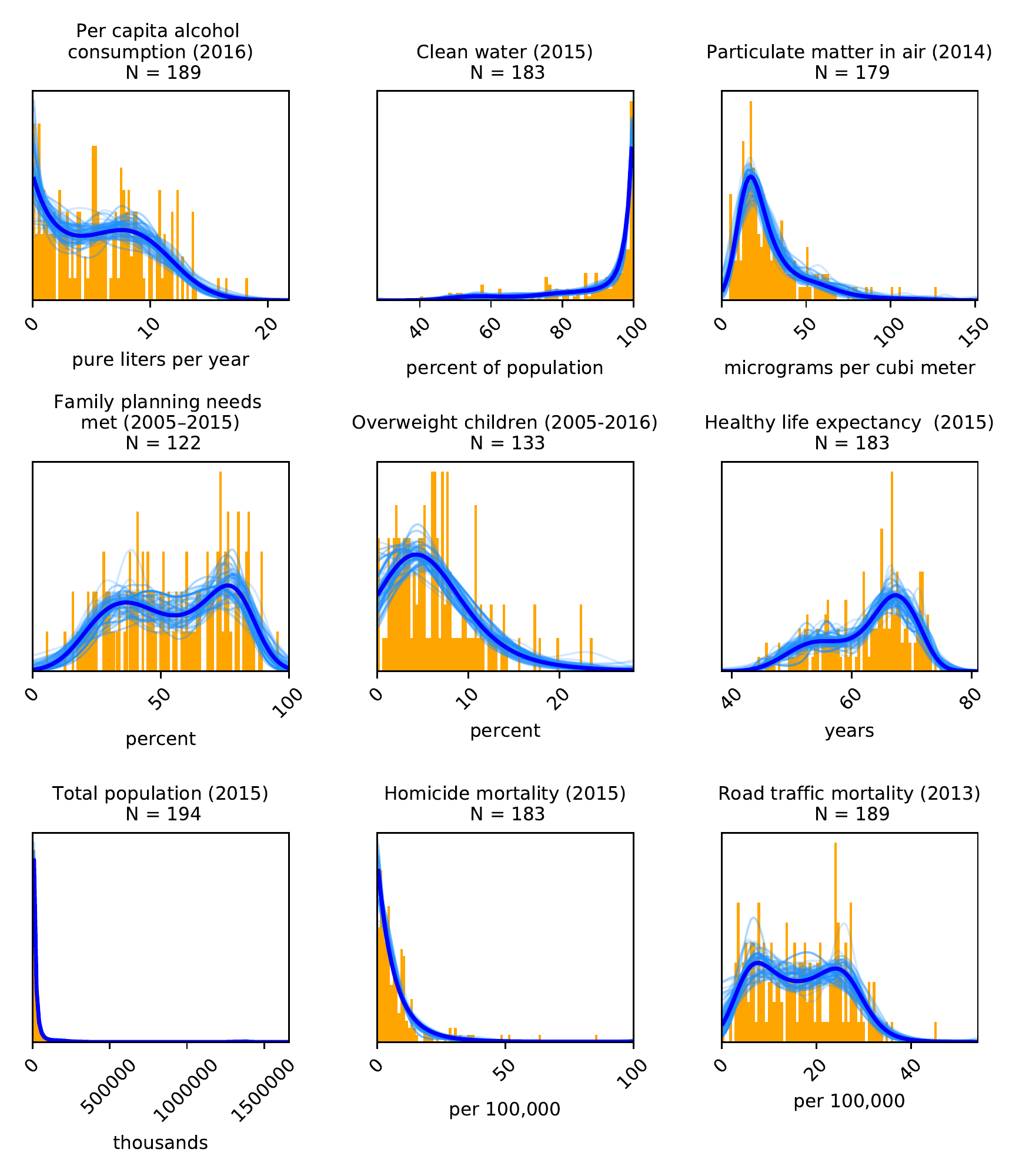}	
\caption{(Color) \textbf{Demonstration of DEFT on data from the World Health Organization (WHO)}. Densities were estimated for 9 different global health indicators reported by the WHO in \cite{WHO:2017ab}. Each datum corresponds to a different country; $N$ varies between panels because of missing data in \cite{WHO:2017ab}. Orange shows a histogram of each global health indicator computed using $G = 100$ grid points. The best DEFT estimate $Q^*$ is shown in dark blue, while 100 posterior-sampled densities $Q \sim p(Q | \data)$ are shown in light blue. As in Fig.\ 3, default DEFT parameters were used for all 9 of these datasets. 
\label{fig:who}} 
\end{figure}

\end{widetext}

\end{document}